\documentclass[aps,prl,twocolumn,superscriptaddress,showpacs]{revtex4}

\usepackage{graphicx}

\RequirePackage{epsfig}

\def\Journal#1#2#3#4{{#1} {\bf #2}, #3 (#4)}
\def\PRL{Phys. Rev. Lett.}
\def\PRA{Phys. Rev. A}

\def\JPB{J. Phys. B: At. Mol. Opt. Phys.}

\begin{document}
\title{Atomic-phase interference devices based on ring-shaped Bose-Einstein condensates: Two ring case}

\author{B. P. Anderson}
\affiliation{Optical Sciences Center, University of Arizona,
Tucson, AZ 85721, USA}

\author{K. Dholakia}
\affiliation{School of Physics \& Astronomy, University of St.
Andrews, North Haugh, St. Andrews, Fife KY16 9SS, Scotland, UK}

\author{E. M. Wright}
\email[]{ewan.wright@optics.arizona.edu}
\affiliation{Optical
Sciences Center, University of Arizona, Tucson, AZ 85721, USA}


\begin{abstract}
We theoretically investigate the ground-state properties and
quantum dynamics of a pair of adjacent ring-shaped Bose-Einstein
condensates that are coupled via tunneling. This device, which is
the analogue of a symmetric superconducting quantum interference
device, is the simplest version of what we term an Atomic-Phase
Interference Device (APHID). The two-ring APHID is shown to be
sensitive to rotation.
\end{abstract}

\pacs{03.75.Fi,03.75.-b}

\maketitle
\section{Introduction}
The last few years have witnessed magnificent advances in the
preparation, manipulation, and exploration of atomic Bose-Einstein
condensates (BECs). These quantum-degenerate systems offer an
excellent experimental platform from which to study a multitude of
nonlinear matter-wave phenomena including four-wave mixing
\cite{phillips}, dark \cite{dark} and bright \cite{bright}
solitons, superfluid vortices \cite{vort}, and the generation and
study of quantized vortices on toroidal atomic traps or rings. In
particular, ring-shaped BECs allow for the study of phenomena
related to persistent currents and rotational motion, with
potential applications to rotation sensing. In this paper, our
goal is to take the first theoretical steps in studying Josephson
coupling between adjacent ring BECs (as opposed to concentric ring
BECs that have been considered previously \cite{TemDevAbr01}). In
particular, we investigate how quantum tunneling between two
condensates trapped in adjacent toroidal traps, formed for example
using optical-dipole traps with Laguerre-Gaussian light beams,
modifies both the ground state properties and quantum dynamics of
the system. The two-ring BEC system is the simplest example of
what we refer to as an Atomic-Phase Interference Device (APHID),
essentially a neutral-atom analog of a SQUID (Superconducting
Quantum Interference Device). The properties of the APHID will be
shown to be strongly influenced by the individual phases of the
matter-waves in the rings.

The remainder of this paper is organized as follows: In the next
section we elucidate the details of the model we use. Following
this, we explore the properties of the ground state and first
excited state of the system. We then look at the Josephson
coupling and the time-dependent solutions, highlighting important
considerations due to the effects of rotation, followed by
concluding remarks.
\section{Basic model}
\begin{figure}
\includegraphics*[width=1.0\columnwidth]{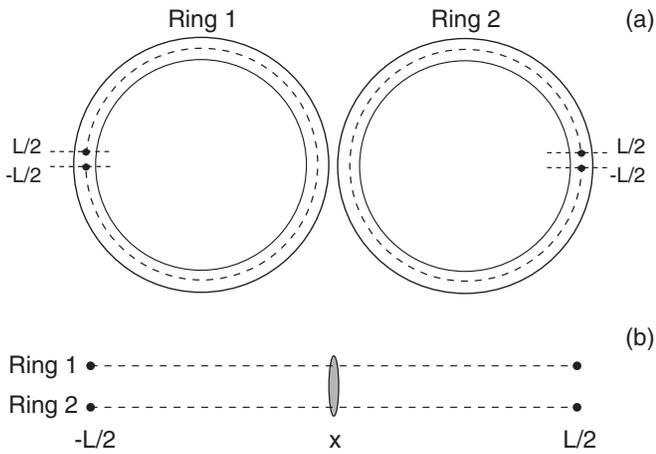}
\caption{(a) The basic model we consider comprises two identical
ring BECs labeled $j=1,2$ which are in close proximity and coupled
via tunnelling, and (b) shows the unfolded rings to which periodic
boundary conditions are applied. The rings come closest together
at the origin $x=0$, where tunneling is represented by a dark
oval.} \label{figure1.RingCoup}
\end{figure}

The basic model we consider is shown in Fig.
\ref{figure1.RingCoup}(a) and comprises two identical ring BECs
labeled $j=1,2$ which are in close proximity, and the whole system
is rotating at an angular frequency $\omega_R$. The close
proximity of the rings allows for spatially dependent tunneling
between them via mode overlap, meaning that the rings are coupled,
allowing Josephson oscillations \cite{TilTil86,Jav86}. Each
individual ring may be realized physically using a toroidal trap
of high aspect ratio $R=L/\ell_{0}$ where $L$ is the toroid
circumference and $\ell_{0}$ the transverse oscillator length
$\ell_{0}=\sqrt{\hbar/m\omega_0}$, with $\omega_0$ the frequency
of transverse oscillations, assumed to be harmonic. The transverse
trap potential is assumed to be symmetric about an axis consisting
of a circle on which the trap potential is minimum. The
longitudinal (circumferential) motion on each ring can be
described approximately by a 1D coordinate $x_j\in [-L/2,L/2]$
obtained by unfolding the ring and applying periodic boundary
conditions, as illustrated in Fig. \ref{figure1.RingCoup}(b).
Then, at zero temperature the quantum dynamics of an atomic BEC
moving on the paired rings may be described by the following
coupled Gross-Pitaevskii equations in a reference frame rotating
at $\omega_R$: \cite{LifPit89,Rok97,JavPaiYoo98,BenRagSme99}
\begin{eqnarray}
i\hbar\frac{\partial\psi_j}{\partial t} & = & \hbar\omega_0\psi_j
- \frac{\hbar^2}{2m}\frac{\partial^2\psi_j}{\partial x^2} -
i(-1)^j\frac{\hbar\omega_R L}{2\pi}\frac{\partial\psi_j}{\partial
x} \nonumber \\
& + & g|\psi_j|^2\psi_j + \hbar\Omega(x)\psi_{3-j}  , \label{GP0}
\end{eqnarray}
where $\psi_j(x,t)$ is the macroscopic wave function for ring
$j=1,2$ with normalization condition
\begin{equation}
\int_0^Ldx\left (|\psi_1(x,t)|^2+|\psi_2(x,t)|^2\right ) = N.
\end{equation}
Here, $N$ is the number of atoms of mass $m$,
$g=4\pi\hbar^2a/(2\pi\ell_0^2m)=2\hbar\omega_0a>0$ is the
effective one-dimensional nonlinear coefficient describing
repulsive many-body interactions, $a$ being the s-wave scattering
length, and $\Omega(x)>0$, which is chosen real and positive, is
the spatially dependent tunneling frequency between the two
rings. In writing Eqs. (\ref{GP0}) we have taken advantage of the
fact that although the atoms in each ring are described by
different coordinates $x_{j=1,2}$, they can nonetheless be
described as moving on the same domain $x\in[-L/2,L/2]$ with the
following caveats: First, the atoms on each ring do not
cross-interact via mean-field effects, and are only coupled via
the spatially dependent tunneling. Second, inspection of Figs.
\ref{figure1.RingCoup}(a) and (b) shows that atoms circulating
from $x=-L/2\rightarrow L/2$ along ring $j=1$ are going
counter-clockwise whereas atoms circulating from
$x=-L/2\rightarrow L/2$ along ring $j=2$ are going clockwise.
This means that although we write the equations using a common
spatial coordinate $x\in [-L/2,L/2]$, propagation in a given $x$
direction corresponds to opposite senses of rotation for the
different rings. This is why the rotation term proportional to
$\omega_R$ in Eq. (\ref{GP0}) has a ring-dependent sign $(-1)^j$.

With reference to Fig. \ref{figure1.RingCoup}(a) we see, for
example, that for an atom moving clockwise from a given reference
point on ring $j=1$, then tunnelling over to ring $j=2$ and moving
counter-clockwise, and finally tunnelling back after orbiting
ring $j=2$ to ring $j=1$ to the original starting point, the atom
crosses the tunnelling region twice. In this sense the coupled
atomic rings are analogous to a symmetric SQUID \cite{TilTil86},
in which two superconducting rings are connected by a weak link,
which has been employed as a magnetometer \cite{Zim70}. The two
ring system, then, is the simplest version of an Atomic-Phase
Interference Device, and we concentrate on the two-ring case in
this paper to explore the basic properties of APHIDs.

The tunneling frequency $\Omega_{max}=\Omega(x=0)$ will be at its
maximum at the point of closest approach of the rings, which we
choose at $x=0$, and will decrease with separation, or
equivalently as $x$ varies away form the origin. Typically, the
tunneling frequency decays exponentially with ring separation.
Thus, $\Omega(x)$ will typically be a bell-shaped function of
$x$, and the spatial extent of the Josephson-coupling will be
much less than the size of the ring $L$. Clearly, for smaller
rings with tighter curvature, $\Omega(x)$ will drop off faster
away from the peak. In the limit $\Omega(x)=0$, Eqs. (\ref{GP0})
reduce to the approximate one-dimensional form previously used to
describe atomic BECs on a toroid.

The conserved N-particle energy functional for the coupled
Gross-Pitaevskii Eqs. (\ref{GP0}) is
\begin{eqnarray}
E & = & N\hbar\omega_0 + \int_0^L dx [ \frac{\hbar^2}{2m}\left (
\left |\frac{\partial\psi_1}{\partial x}\right |^2 + \left
|\frac{\partial\psi_2}{\partial x}\right |^2 \right ) \nonumber \\
& + & i\frac{\hbar\omega_R L}{2\pi} \left (
\psi_2^*\frac{\partial\psi_1}{\partial x} -
\psi_1^*\frac{\partial\psi_2}{\partial x} \right ) \nonumber \\
& + & \frac{g}{2}\left ( |\psi_1|^4 + |\psi_2|^4 \right ) +
\hbar\Omega (x)\left (\psi_1\psi_2^* + \psi_1^*\psi_2 \right ) ]
, \label{EN}
\end{eqnarray}
giving the energy per particle $\epsilon=E/N$. Since in this
paper the transverse confinement energy $\hbar\omega_0$ is
assumed the same for both rings and simply redefines the zero of
energy, we hereafter drop this energy term for simplicity in
notation.
\section{Ground and first excited states}
In this section we examine the properties of the ground and first
excited states of a non-rotating ($\omega_R=0$) pair of coupled
ring BECs using a simple model to expose the main features.
\subsection{Zero-coupling limit}
It is useful in assessing the ground state properties to consider
the non-coupled case with $\Omega(x)=0$. If all $N$ atoms are
homogeneously distributed on just one of the rings, with
$\psi_j=\sqrt{N/L}$ and $\psi_{3-j}=0$, then according to Eq.
(\ref{EN}) the energy per particle is $\epsilon_{trap} = gn/2$,
where $n=N/L$ is the linear atomic density. In contrast, when the
atoms are equally split between the two rings, but still
homogeneously distributed on each ring, $|\psi_j|=\sqrt{N/2L}$,
and the energy per particle is
\begin{equation}
\epsilon_{1/2} = \frac{gn}{4}  , \label{esplit}
\end{equation}
irrespective of the relative phase between the macroscopic wave
functions of the two rings. Energetically speaking then, in the
absence of coupling the lowest-energy state is that in which the
atoms are equally split between the rings as this minimizes the
mean-field energy.
\subsection{Coupled solutions}
To proceed we now re-introduce the coupling and look for solutions
where the atoms are equally split between the rings. In
particular we consider solutions where the macroscopic wave
functions of the two rings are in-phase $(+)$ and out-of-phase
$(-)$ by making the ansatz
\begin{equation}
\psi_j(x,t) = \frac{(\pm 1)^j}{\sqrt{2}}e^{-i\mu_\pm
t/\hbar}\varphi_\pm(x) , \qquad j=1,2\label{ansatz}
\end{equation}
with $\varphi_\pm(x)$ the mode profiles on each ring and $\mu_\pm$
the corresponding chemical potentials. Then substituting in Eq.
(\ref{GP0}) we obtain
\begin{equation}
\mu_\pm\varphi_\pm =
-\frac{\hbar^2}{2m}\frac{d^2\varphi_\pm}{dx^2} +
\frac{g}{2}|\varphi_\pm|^2\varphi_\pm \pm
\hbar\Omega(x)\varphi_\pm , \label{GP1}
\end{equation}
and $\int dx|\varphi_\pm(x)|^2=N$. On general grounds, the out-of
phase $(-)$ solution corresponds to the ground state. This can be
seen from Eq. (\ref{GP1}) where the spatially dependent coupling
$\Omega(x)>0$, which is typically bell-shaped, plays the role of a
confining (de-confining) potential for the out-of-phase (in-phase)
solution, thereby allowing for lower energy in comparison to the
case without coupling.

In the limit $\Omega=0$ Eq. (\ref{GP1}) also has the well-known
dark soliton solution
\cite{Reinhardt,Dum,Scott,Jackson,Muryshev,CarClaRei00} on the
infinite domain $L\rightarrow\infty$
\begin{equation}
\varphi_\pm(x) = \varphi_0(x)= \sqrt{n}\tanh\left
(\frac{(x+x_0)}{\sqrt{2}x_h} \right ) , \label{darksol}
\end{equation}
with $\mu_0 = gn/2$, where $n$ is the linear density of the
background (in the thermodynamic limit, where
$N\rightarrow\infty$ and $L\rightarrow\infty$, $N/L\rightarrow n$
remains non-zero). The healing length $x_h$ is derived from the
relation
\begin{equation}
\frac{\hbar^2}{2mx_h^2} = \frac{gn}{2} .  \label{xh}
\end{equation}
The dark soliton solution represents a flat background density
profile with a hole of width $x_h<<L$ located $x=-x_0$, at which
location a phase jump of $\pi$ also occurs as $\varphi_0$ goes
through zero. In the thermodynamic limit the energy per particle
associated with the dark soliton solution calculated using Eq.
(\ref{EN}) is $\epsilon_0=ng/4=\epsilon_{1/2}$, that is, it is the
same as that in Eq. (\ref{esplit}) for a homogeneous density on
each ring without coupling. This arises because in the
thermodynamic limit $x_h/L\rightarrow 0$, meaning that any energy
increase due to the hole in the density makes a negligible effect
on average; in other words, the hole in the density occupies a
vanishingly small portion of the ring.
\subsection{Analytic approximation}
\begin{figure}
\includegraphics*[width=1.0\columnwidth]{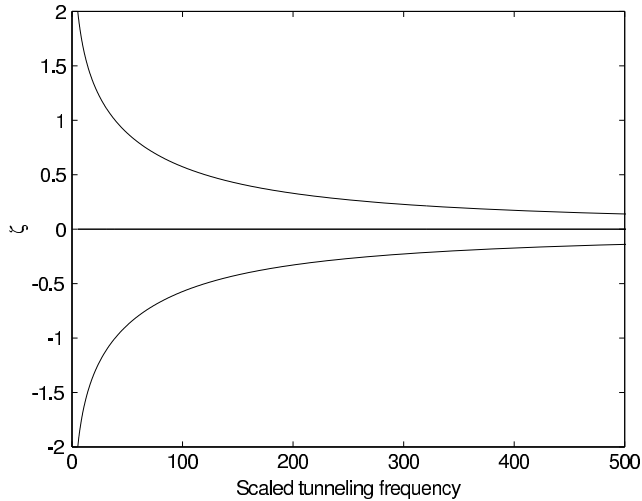}
\caption{Plot of $\zeta$ versus $\hbar\Omega_{max}\cdot d/g$ for
$n/n_s=10^4$, with $\zeta=x_+/\sqrt{2}x_h, \zeta>0$ and
$\zeta=x_-/\sqrt{2}x_h, \zeta<0$.} \label{figure2.RingCoup}
\end{figure}

\begin{figure}
\includegraphics*[width=1.0\columnwidth]{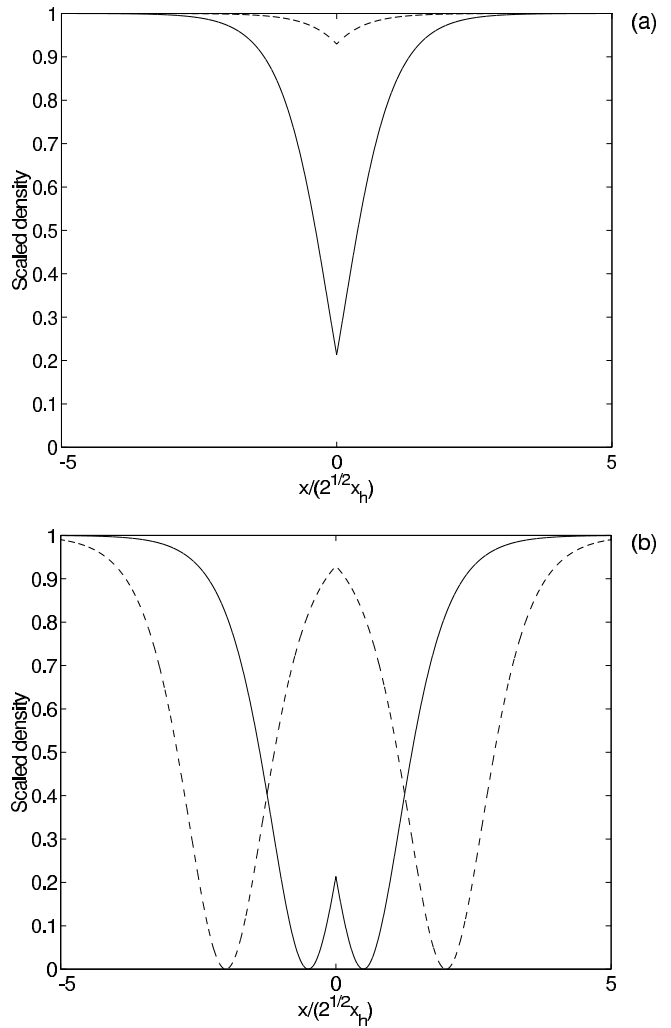}
\caption{Scaled density profiles $|\varphi_\pm|^2/n$ for
$n/n_s=10^4$, $\hbar\Omega_{max}\cdot d/g=95, \zeta=0.5$ (solid
lines), $\hbar\Omega_{max}\cdot d/g=6.3, \zeta=2$ (dashed lines)
and (a) the out-of-phase or ground state solution, and (b) the
in-phase solution.} \label{figure3.RingCoup}
\end{figure}

In general, numerical methods are required to solve Eq.
(\ref{GP1}) for given parameters and tunneling profile
$\Omega(x)$. In order to obtain insight into the ground-state
properties, we employ a simple model
\begin{equation}
\Omega(x) = \Omega_{max}\cdot d\cdot\delta(x)  , \label{delta}
\end{equation}
where $\Omega_{max}$ is the maximum tunneling frequency and $d$
is the length of the tunneling region. This delta-function
approximation will apply when $d$ is much less than any other
characteristic length scale of the problem, namely the ring length
$L$ and the healing length $x_h$. For the stationary coupled-ring
solutions described by Eq. (\ref{GP1}), where $\Omega(x)$ plays
the role of a single-particle potential, the delta-function
approximation yields a quantum-contact interaction
\cite{TruWan00}. Substituting Eq. (\ref{delta}) in (\ref{GP1}) and
integrating from $x=0_-$ to $x=0_+$ across the junction, we find
that the action of the delta-function coupling is equivalent to a
condition on the macroscopic wave function derivative
\begin{equation}
\frac{\hbar^2}{2m}\left (\frac{d\varphi_\pm}{dx}|_{x=0_+} -
\frac{d\varphi_\pm}{dx}|_{x=0_-} \right ) = \pm
\hbar\Omega_{max}\cdot d\cdot\varphi_\pm(0)  . \label{boundcond}
\end{equation}
In the limit $L>>x_h>>d$ we further impose the condition that
$\varphi_\pm(x)$ is symmetric around $x=0$ in order to satisfy the
periodic ring boundary conditions, and we approximate
\begin{equation}
\varphi_\pm(x)\approx \sqrt{n}\tanh\left
(\frac{(x+x_\pm)}{\sqrt{2}x_h} \right ) , \qquad x>0 .
\label{approx}
\end{equation}
With this approximation there is a cusp in $\varphi_\pm(x)$ at
x=0, and the solution is extended to $x<0$ by imposing reflection
symmetry around the origin.  We can solve for the variables
$x_\pm$ by substituting the approximate solution (\ref{approx}) in
the boundary condition (\ref{boundcond}), which yields
\begin{equation}
\pm\hbar\Omega_{max}\cdot d = \left (\frac{\hbar^2}{m\sqrt{2}x_h}
\right ) \frac{[1-\tanh ^2(x_\pm/\sqrt{2}x_h)] } {
\tanh(x_\pm/\sqrt{2}x_h) } .
\end{equation}
Since $\Omega_{max}>0$ we find by inspection that the in-phase
solutions correspond to $x_+>0$ and the out-of-phase solutions to
$x_-<0$. By introducing a dimensionless  parameter
$\zeta=x_+/\sqrt{2}x_h$, with $\zeta>0$ and
$\zeta=x_-/\sqrt{2}x_h$, with $\zeta<0$, and using Eq. (\ref{xh})
for the healing length, we may write the above equation as
\begin{equation}
\frac{\hbar\Omega_{max}\cdot d}{g} = \sqrt{\frac{n}{2n_s}}
\frac{[1-\tanh ^2(\zeta)] } { \tanh(|\zeta|) } , \label{sol1}
\end{equation}
where $n_s=mg/\hbar^2$ is a scaled density. Figure
\ref{figure2.RingCoup} shows a plot $\zeta$ versus the scaled
tunneling frequency $\hbar\Omega_{max}\cdot d/g$ for
$n/n_s=10^4$.  Figure \ref{figure3.RingCoup} shows examples of
scaled density profiles $|\varphi_\pm|^2/n$ for
$\hbar\Omega_{max}\cdot d/g=95, \zeta=0.5$ (solid lines),
$\hbar\Omega_{max}\cdot d/g=6.3, \zeta=2$ (dashed lines) and (a)
the out-of-phase or ground-state solution, and (b) the in-phase
solution.  Density cusps in the solutions are evident, though we
note that the ground-state density does not extend down to zero.
The key features of the ground state are that as the scaled
tunneling frequency $\hbar\Omega_{max}\cdot d/g$ is increased the
depth of the density profile increases, the density at the origin
going to zero as $\hbar\Omega_{max}\cdot d/g\rightarrow\infty$,
and the width of the density hole also increases, approaching
$x_h$ as $\hbar\Omega_{max}\cdot d/g\rightarrow\infty$. The
in-phase solution is different in that it displays two density
zeros and an on-axis maximum that is a cusp, shown in Fig.
\ref{figure3.RingCoup}(b). Furthermore, inspection of the
in-phase solution shows that its sign reverses through each
density zero, and there are two sign reversals around each ring
to ensure that the wave functions are single-valued. The in-phase
solution therefore has a phase structure like a pair of dark
solitons on each ring. For small $\hbar\Omega_{max}\cdot d/g<<1$,
the density zeros are far apart (dashed line in Fig.
\ref{figure3.RingCoup}(b) for $\hbar\Omega_{max}\cdot d/g=0.2,
\zeta=2$), but come together at the origin as
$\hbar\Omega_{max}\cdot d/g\rightarrow\infty$ (solid line in Fig.
\ref{figure3.RingCoup}(b) for $\hbar\Omega_{max}\cdot d/g=3,
\zeta=0.5$). Thus, for both the in-phase and out-of-phase
solutions the density vanishes at the origin as
$\hbar\Omega_{max}\cdot d/g\rightarrow\infty$, and we have
\begin{equation}
\varphi_\pm(x)\approx \sqrt{n}\tanh\left (\frac{x}{\sqrt{2}x_h}
\right ) .
\end{equation}

\begin{figure}
\includegraphics*[width=1.0\columnwidth]{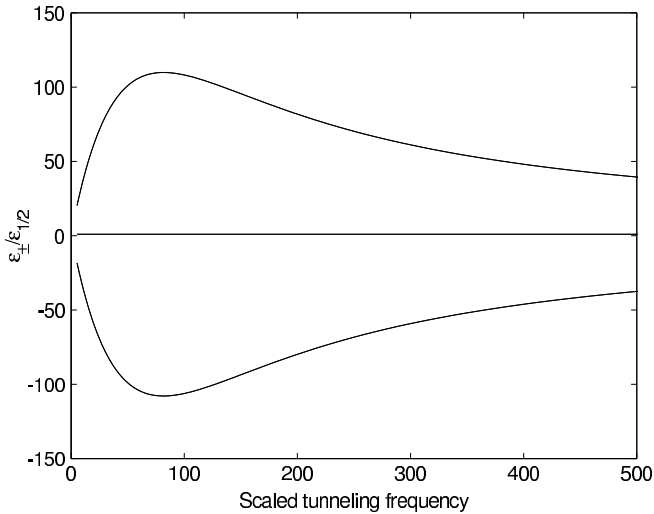}
\caption{Scaled energy per particle $\epsilon_\pm/\epsilon_{1/2}$
versus $\hbar\Omega_{max}\cdot d/g$ for $n/n_s=10^4$, the upper
solid line corresponding to the in-phase $(+)$ solution and the
lower solid line to the out-of-phase $(-)$ or ground-state
solution.} \label{figure4.RingCoup}
\end{figure}

A quantity of physical interest here is the energy per particle
$\epsilon_\pm$ for the two solutions. Using the above approximate
solution in the energy functional (\ref{EN}) we find in the
thermodynamic limit
\begin{equation}
\epsilon_\pm = \frac{ng}{4} \pm \hbar\Omega_{max}\cdot d\cdot n
\tanh^2(|\zeta|) , \label{sol2}
\end{equation}
where the solution is again parameterized by $\zeta$. Note that in
the limit of zero coupling $\Omega_{max}\rightarrow 0$, the
energies per particle of the two solutions become the same and
equal to that of the equally split solution $\epsilon_{1/2}=ng/4$
as they should. Using Eq. (\ref{sol1}) in (\ref{sol2}) we obtain
finally
\begin{equation}
\epsilon_\pm = \epsilon_{1/2} \left (1 \pm
\sqrt{\frac{8n}{n_s}}\tanh(|\zeta|)[1-\tanh ^2(\zeta)] \right ) ,
\label{sol3}
\end{equation}
which is once again parameterized by $\zeta$. Figure
\ref{figure4.RingCoup} shows $\epsilon_\pm/\epsilon_{1/2}$ versus
$\hbar\Omega_{max}\cdot d/g$ for $n/n_s=10$, the upper solid line
corresponding to the in-phase $(+)$ solution and the lower solid
line to the out-of--phase $(-)$ or ground state solution. For
small values of the scaled tunneling frequency
$\hbar\Omega_{max}\cdot d/g<1$ the energy per particle for the
in-phase (out-of-phase) solution initially increases (decreases)
away from $\epsilon_{1/2}$ for zero-coupling, and this is expected
physically. However, as the scaled tunneling frequency is
increased further the energy per particle for the in-phase
(out-of-phase) solution reaches a turning point at
$\hbar\Omega_{max}\cdot d/g\approx 2$, then decreases (increases),
and both $\epsilon_\pm$ tend back to the zero-coupling  value
$\epsilon_{1/2}$ for $\hbar\Omega_{max}\cdot
d/g\rightarrow\infty$. The reason for this is that, as discussed
above, for both solutions the density tends to zero at the origin
$x=0$ where the junction is concentrated in the limit
$\hbar\Omega_{max}\cdot d/g\rightarrow\infty$, so the Josephson
coupling is rendered inoperative and the energy per particle tends
to that for zero-coupling.
\section{Time-dependent solutions}
\subsection{Scaled equations}
For purposes of numerical simulations we introduce a simple
Gaussian model for the spatially dependent Josephson-coupling
\begin{equation}
\Omega(x) = \Omega_{max}e^{-x^2/w^2}=\Omega_{max}\cdot d\cdot f(x)
, \label{Omega}
\end{equation}
with $\Omega_{max}$ the maximum tunneling frequency and $w<<L$
the width of the coupling region around $x=0$. We also introduce
the normalized Gaussian $f(x)=\exp(-x^2/w^2)/\sqrt{\pi w^2}$ for
which $d=\sqrt{\pi w^2}$ so that effective parameters can be
compared with the previous section. Then introducing the scaled
variables
\begin{equation}
\tau=t\cdot(ng/\hbar),\quad \xi=x/L,\quad \psi_j=\sqrt{n}\phi_j,
\label{scale}
\end{equation}
with $n=N/L$ the mean density as before, we obtain with $j = 1,2$,
\begin{eqnarray}
i\frac{\partial\phi_j}{\partial\tau} & = &
-\frac{\beta}{2}\frac{\partial^2\phi_j}{\partial \xi^2} -
i(-1)^j\left (\frac{\nu}{2\pi}\right )
\frac{\partial\phi_j}{\partial \xi} \nonumber \\
& & + |\phi_j|^2\phi_j + \eta e^{-\xi^2/\Delta^2}\psi_{3-j},
 \label{GPS}
 \end{eqnarray}
where $\int d\xi[|\phi_1|^2+|\phi_2|^2]=1$, and
\begin{equation}
\Delta=\frac{w}{L}<<1,\quad
\eta=\frac{\hbar\Omega_{max}}{ng},\quad\nu=\frac{\hbar\omega_R}{ng},
\quad \beta = \frac{(n/n_s)}{N^2} . \label{defs}
\end{equation}
These are the scaled equations used for our numerical study. We
have solved the equations numerically using the split-step
Fast-Fourier transform method \cite{FleMorFei76}.

To study the quantum dynamics of coupled-ring BECs we shall use an
initial condition at $\tau=0$ where all $N$ atoms are on one ring
in a vortex state of winding number $p$. This may be realized, for
example, by condensing the atoms on one ring in the absence of the
other, stirring the BEC to create the vortex \cite{BraRei01}, and
then turning on the second ring. Sauer {\it et al.}
\cite{SauBarCha01} have demonstrated a 2-cm diameter magnetic
storage ring for laser-cooled, and Arnold and Riis
\cite{ArnRii02} are working towards realizing a 10 cm diameter
magnetically trapped toroidal BEC. One scheme for turning rings
off and on is to use toroidal optical dipole traps
\cite{StaAndChi98} formed by Laguerre-Gaussian beams piercing a
two-dimensional BEC to create the rings
\cite{SalParRea99,WriArlDho00,TemDevAbr01}, or alternatively
using scanned laser beams to form the toroidal traps
\cite{Dur99}. Cavity field enhancement may also be used to allow
for large-radius toroidal traps \cite{FreDho02}. Regardless of
experimental method, the initial condition we take is
\begin{equation}
\phi_1(\xi,0) = e^{2\pi ip\xi} , \quad \phi_2 = 0  . \label{init}
\end{equation}
\subsection{Resonance conditions}
To proceed we examine the resonance conditions leading to the
initial exchange of atoms from ring $1\rightarrow 2$ using
first-order perturbation theory. For the initial condition
(\ref{init}) we choose the zeroth-order solution as that for
$\nu=0$
\begin{equation}
\phi_1^{(0)}(\xi,\tau)= e^{2\pi ip\xi} e^{-i(2\pi^2\beta
p^2-p\nu+1)\tau}  . \label{zeroth}
\end{equation}
Then writing the first-order solution for ring 2 in the form
\begin{equation}
\phi_2^{(1)}(\xi,\tau)= \sum_{q=-\infty}^\infty a_q (\tau)e^{2\pi
iq\xi} e^{-i(2\pi^2\beta q^2+q\nu)\tau}  ,
\end{equation}
yields
\begin{equation}
|a_q(\tau)|^2=4\eta^2{\cal F}_{pq}^2\cdot
 \frac{\sin^2(\chi_{pq}\tau/2)}{\chi_{pq}^2} ,
\end{equation}
where
\begin{eqnarray}
{\cal F}_{pq} & = & \sqrt{\pi}\Delta e^{-\pi^2\Delta^2(p-q)^2} , \nonumber \\
\chi_{pq} & = & 2\pi^2\beta(p^2-q^2)-\nu(p+q)+1  .
\end{eqnarray}
The vortex states $q$ of ring 2 are therefore excited and
generally exhibit small oscillations except at resonance where
$\chi_{pq}$ becomes small. The level of excitation of the $q^{th}$
vortex state is also dictated by the factor ${\cal F}_{pq}$, but
since we assume a narrow junction $w/L=\Delta<<1$, this factor
allows for almost constant excitation ${\cal F}_{pq}\approx
\sqrt{\pi}\Delta$ in the range $q=p\pm\delta q$ with
\begin{equation}
\delta q = \frac{1}{\pi\Delta} >> 1.
\end{equation}

Consider first the case that the system is not rotating $\nu=0$:
Resonance occurs for that integer value of $q_r$ for which
$\chi_{pq}$ is equal to or closest to zero
\begin{equation}
q_r^2 = p^2 + \frac{1}{2\pi^2\beta} , \label{qr}
\end{equation}
the width of the resonance being
\begin{equation}
\Delta q\approx \frac{1}{2\pi^2(p+q_r)\beta} . \label{Dq}
\end{equation}
When the width of the resonance is small $\Delta q<1$ the initial
vortex of index $p$ in ring 2 will selectively couple to vortices
with mode indices $q_r$ satisfying Eq. (\ref{qr}) in ring 2,
giving rise to relatively simple few mode dynamics. In contrast,
when $\Delta q>>1$ the initial vortex of index $p$ in ring 2 will
couple to a broad range of vortices with mode indices
$q_r\pm\Delta q$ in ring 2, giving rise to multi-mode dynamics and
complex behavior. In addition, for Josephson oscillations to
occur the tunneling energy per particle averaged over the ring
length $(1/L)\int
dx\cdot\hbar\Omega(x)=\hbar\Omega_{max}\sqrt{\pi}w/L$ should be
for the same order as the mean-field energy per particle $ng$, or
\begin{equation}
\eta =\frac{\hbar\Omega_{max}}{ng} \sim \frac{1}{\sqrt{\pi}\Delta}
.
\end{equation}
This gives an estimate of the scaled tunneling frequency $\eta$
to obtain Josephson oscillations.
\subsection{Numerical results}
\begin{figure}
\includegraphics*[width=1.0\columnwidth]{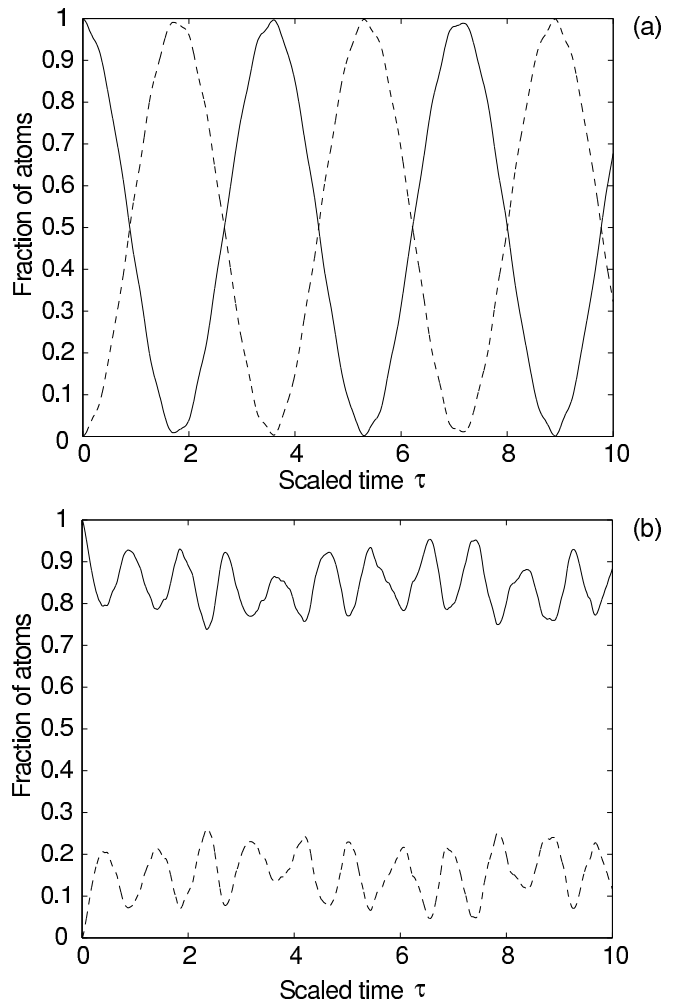}
\caption{Fraction of atoms in each ring for
$\Delta=10^{-2},\eta=50$, and (a) $\beta=1, \Delta q=0.22$, and
(b) $\beta=5.1\times 10^{-2}, \Delta q=1$.}
\label{figure5.RingCoup}
\end{figure}

\begin{figure}
\includegraphics*[width=1.0\columnwidth]{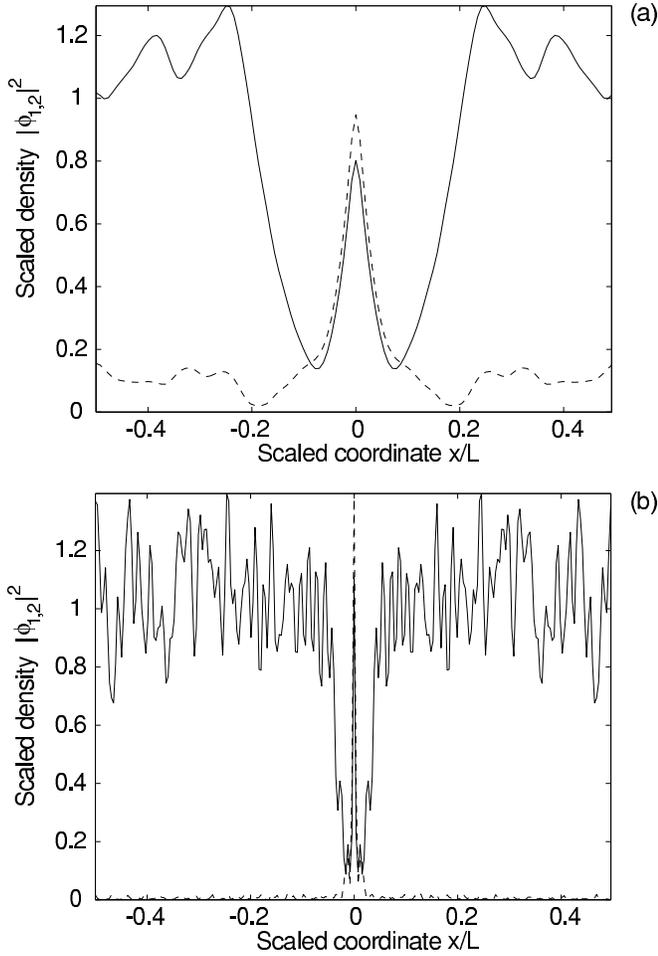}
\caption{(a) Spatial density in ring $j=1$ (solid line) and $j=2$
(dashed line) for the same parameters as Fig.
\ref{figure5.RingCoup}(b) with $\tau=10$, $\beta=5.1\times
10^{-2}$, and $\Delta q=1$; (b) spatial density in ring $j=1$
(solid line) and $j=2$ (dashed line) for $\beta=5.1\times
10^{-4}, \Delta q=10$ for $\tau=10$.} \label{figure6.RingCoup}
\end{figure}

Here we present some examples of the dynamics of coupled ring
BECs. For all the simulations we set $p=0,\Delta=10^{-2}$, and
$\eta=50$. Consider first that the initial state corresponds to
the ground state ($p=0$) of ring 1. From Eqs. (\ref{qr}) and
(\ref{Dq}) we obtain
\begin{equation}
q_r = \sqrt{\frac{1}{2\pi^2\beta}} = \Delta q ,
\end{equation}
that is, the width of the resonance $\Delta q$ is equal to the
resonant value $q=q_r$. Figure \ref{figure5.RingCoup}(a) shows the
fraction of atoms in each ring for $\beta=1$ for which $q_r=\Delta
q =0.22$, and complete Josephson oscillations between the two
rings are evident. In this case the density profiles in the two
rings are largely flat as resonant coupling occurs between
$p=0,q_r\approx 0$. In contrast, for $\beta=5.1\times 10^{-2}$ as
shown in Fig. \ref{figure5.RingCoup}(b) for which $q_r=\Delta q
=1$, the Josephson oscillations are now incomplete. Physically,
there are multiple modes involved in ring $j=2$ with $q=0,\pm
1,\pm 2$, and the resulting multi-mode dynamics is what
frustrates the Josephson oscillations for $\Delta q\ge 1$. The
multi-mode dynamics manifests itself as spatial density
modulations in the two rings as shown in Fig.
\ref{figure6.RingCoup}(a) for the same parameters as in Fig.
\ref{figure5.RingCoup}(b) and $\tau=10$. For even lower density
$\beta=5.1\times 10^{-4}$ for which $q_r=\Delta q =10$, the
Josephson oscillations are all but extinguished, and the spatial
density profiles in rings $j=1,2$ are shown in Figs.
\ref{figure6.RingCoup}(b) for $\tau=10$. Clearly, the multi-mode
nature of the solution allows the coupling due to tunneling to
concentrate around the coupling region, hence reducing the net
fraction of atoms transferred between the rings.


Some estimates of parameters are in order. Using $g=2\hbar\omega_0
a$ gives $ng=2\hbar\omega_0 N(a/L)$, and
$\Omega_{max}=2\eta\omega_0 N(a/L)$. Then for $\omega_0=2\pi\times
10^2$ rad$\cdot s^{-1}, N=10^3, L=1$ cm, $a=5$ nm, we find
$\Omega_{max}=2\pi\times 5$ rad$s^{-1}$, and $\tau$ is time in
units of $\hbar/ng=1.6$ s, so the Josephson oscillations in Fig.
\ref{figure5.RingCoup}(a) occur on a time scale of seconds.
Setting $m=10^{-25}$ kg we obtain $n_s=mg/\hbar^2\approx 63$
cm$^{-1}$, and for $n=N/L=10^3$ cm$^{-1}, \beta = 1.6\times
10^{-5}$. It is important that $n/n_s>1$ to ensure that the
one-dimensional gas acts as a BEC as opposed to a Tonks gas
\cite{Ols98,PetShlWal00}. The parameter $\beta=(n/n_s)/N^2$ is
proportional to $1/\omega_0$ and $1/N$ so we can increase $\beta$
by decreasing either the number of atoms and/or the transverse
oscillator frequency with respect to the above values.
\begin{figure}
\includegraphics*[width=1.0\columnwidth]{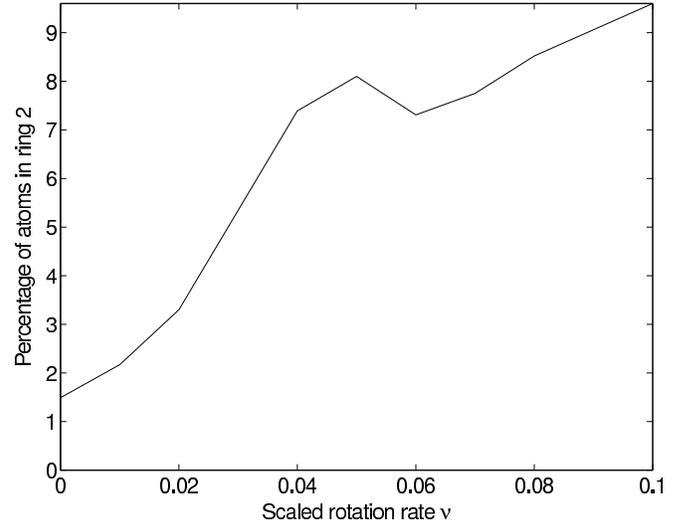}
\caption{Percentage of atoms in ring $j=2$ as a function of scaled
rotation rate $\nu=\hbar\omega_R/ng$.} \label{figure7.RingCoup}
\end{figure}
\subsection{Effects of rotation}
An interesting feature of the two ring APHID is that the condition
$\chi_{pq}=2\pi^2\beta(p^2-q^2)-\nu(p+q)+1\rightarrow 0$ for
resonant coupling between the rings is dependent on the scaled
rotation rate $\nu=\hbar\omega_R/ng$. In particular we find for
$p=0$
\begin{equation}
q_r = \frac{1}{4\pi^2\beta}\left [ -\nu \pm
\sqrt{\nu^2+8\pi^2\beta} \right ] . \label{qrnu}
\end{equation}
This implies that for scaled rotation rates
$|\nu|>\sqrt{8\pi^2\beta}$ the rotation of the entire APHID will
affect the coupling. Consider then a case where without rotation
$\Delta q>>1$ so that the Josephson-oscillations are all but
extinguished and the atoms remain on ring $1$. Then as the scaled
rotation rate $\nu$ is increased from zero, inspection shows that
one solution $q_r$ in Eq. (\ref{qrnu}) moves towards resonance
while the other moves further away. Therefore, starting from a
detuned case with minimal coupling, increased rotation leads to
increased coupling which can then be detected via the number of
atoms on ring $2$ at a fixed detection time. Figure
\ref{figure7.RingCoup} shows the percentage of the atoms in ring
$2$ versus scaled rotation rate $\nu$ at time $\tau=10$ and
$\Delta = 0.01, \kappa=50,\beta=5.1\times 10^{-4}$, for which
$q_r=\Delta q =10$, and the effect of rotation dependent coupling
between the rings is clearly exhibited. Some points are worth
making here: First, the rotation causes the number of atoms in
ring $2$ to change by about 10\% of the total number of atoms, so
experimentally it will be necessary to control the initial number
of atoms on ring $1$ to better than this percentage. Furthermore,
it would be a challenge to detect the small number of atoms in
ring $2$. Second, for our particular example with $p=0$ the
number of atoms in ring $2$ is sensitive to the magnitude but not
the sign of the rotation, but this can be changed by having $p\ne
0$ in which case the the coupling becomes sensitive to the sign
of $\nu$. Third, the sensitivity of the atom number to rotation
rate increases with the observation time $\tau$ chosen,
remembering that we are in a far-off-resonant situation so
coupling happens slowly. Finally, the number of atoms in ring $2$
is not necessarily a monotonic function of the rotation rate, as
seen from Fig. \ref{figure7.RingCoup}, which will limit the range
of rotation rates that can be uniquely measured. Nonetheless, we
feel this is an interesting phenomena which may have utility for
rotation sensing with further development.

To gain some sense of the sensitivity of this scheme we use the
same parameters as the previous section for which $\tau$ is time
in units of $\hbar/ng=1.6$ s. Then a value of $\nu=0.01$
corresponds to a rotation rate $\omega_R=2\nu\omega_0
N(a/L)=2\pi\times 10^{-3}$ rad$s^{-1}$ which is one hundred times
higher than the Earth's rotation rate at the poles. However, if we
are willing to reduce the transverse oscillator frequency to
$\omega_0=2\pi\times 1$ rad$s^{-1}$, then $\nu=0.01$ corresponds
to the Earth's rotation rate, but then time is in units of $160$ s
in the figures! We are currently working on schemes involving
multiple-ring APHIDs to enhance the rotation sensitivity.
\section{Summary and conclusions}
In summary, we have presented a theoretical investigation of a
pair of ring BECs coupled by tunneling as the simplest example of
a potential Atomic Phase Interference Device. We have shown that
the two-ring APHID has interesting ground-state properties, with
density profiles reminiscent of dark soliton states around the
point of contact of the rings. Furthermore, we have demonstrated
that Josephson oscillations between the two rings can occur, and
that these oscillations are sensitive to the state of rotation of
the APHID. In particular, if all the atoms are prepared on one
ring, then the number of atoms transferred to the second ring in
a given time span is a measure of the rotation rate of the APHID.
Although the two-ring APHID was found to be not very rotation
sensitive, we believe APHIDs are worthy of further study as
multi-ring APHIDs will display enhanced sensitivity to the
relative phase between the rings hence potentially leading to
increased rotation sensitivity. We shall be exploring multi-ring
APHIDs in future research. \vspace{0.2cm}

\noindent This work was supported by the Office of Naval Research
Contract No. N00014-99-1-0806, the U.S. Army Research Office, and
the Royal Society of Edinburgh. KD acknowledges the support of
the UK Engineering and Physical Sciences Research Council.

\end{document}